\documentstyle[epsfig,12pt]{article}
\catcode`@=11

\renewcommand\section{\@startsection {section}{1}{\z@}%
{-3.5ex \@plus -1ex \@minus -.2ex}%
{0.3ex \@plus.2ex}%
{\normalfont\normalsize\bfseries}}

\def\@fnsymbol#1{\ifcase#1\or *\or \dagger\or \ddagger\or
\mathchar "278\or **\or \dagger\dagger \or \ddagger\ddagger \or
\mathchar "278 \mathchar "278\or
***\or\dagger\dagger\dagger\or\ddagger\ddagger\ddagger\or \mathchar
"278 \mathchar "278 \mathchar "278 \else\@ctrerr\fi\relax}

\def\fnsymbol#1{\@fnsymbol{\@nameuse{c@#1}}}
\def\@fnsymbol#1{\ifcase#1\or *\or ** \or *** \or
**** \or ***** \or \******\or *******\or \dagger\dagger \or
\ddagger\ddagger \else\@ctrerr\fi\relax}
\catcode`@=12

\newcommand{\myskip}{\vspace{0.95\baselineskip}}
\newcommand{\mysection}[1]{\par\myskip\noindent\textbf{#1}\myskip\par}
\begin{document}
\begin{center}
{\large\textbf{Effects of Spin Polarization on Electron Transport
in Modulation Doped Cd$_{1-x}$Mn$_{x}$Te/Cd$_{1-y}$Mg$_{y}$Te:I
Heterostructures }}

\myskip T.\ Andrearczyk, J.\ Jaroszy\'nski,  G.\ Karczewski, J.\
Wr\'obel, T.\ Wojtowicz, and T.\ Dietl\\ Institute of Physics,
Polish Academy of Sciences, al. Lotnik\'ow 32/46, PL 02--668
Warszawa, Poland

\myskip E.\ Papis, E. Kami\'nska, and A.\ Piotrowska\\ Institute of
Electron Technology, al. Lotnik\'ow 32/46, PL 02--668 Warszawa,
Poland \mysection{Abstract}

\myskip
\parbox{4.5in}{We examine and identify magnetoresistance mechanisms in
2D system containing a sizable concentration of magnetic ions. We
argue that some of these mechanisms can serve as a tool to measure
spin polarization. Lack of spin degeneracy and enhanced
localization make it possible to detect an additional QHE plateau
associated with extended states floating-up in vanishing magnetic
field.}
\end{center}

In diluted magnetic semiconductors (DMS), owing to the strong s-d
exchange between the effective mass carriers and the localized Mn
spins, a high degree of spin-polarization of the carrier liquid can
be achieved in moderately strong magnetic fields \cite{Diet94}.
This offers a tool to find out how properties of quantum structures
evolve as a function of spin polarization. Our work extends
previous transport studies of II-VI DMS low-dimensional systems
\cite{Grab93,Jaro95,Smor97,Jaro98,Sama98,Jaro00} in two major
directions. First, one of the challenging problems that face
spintronics \cite{Prin98} is how to read the local spin
configuration. We have combine millikelvin studies of low-field
magnetoconductance with  theoretical calculations in order to
single out possible effects of spin-polarization on electron
transport in high electron mobility  modulation-doped
heterostructures of Cd$_{1-x}$Mn$_x$Te/Cd$_{1-y}$Mg$_y$Te:I. Based
on our results, we argue that the temperature dependence of
conductivity can serve as a tool to detect and quantify the degree
of carrier polarization generated, for instance, by spin injection.

Second, it has been already demonstrated that the large
spin-splitting in Cd$_{1-x}$Mn$_x$Te makes it possible to test
quantitatively scaling theory of the quantum Hall effect (QHE) over
a wide range of the filling factors $\nu$ \cite{Jaro00}.
Importantly, such a system offers also a novel environment to
examine the transition between electron liquid and quantum Hall
states. Scaling theory of localization predicts that only localized
states exist at zero magnetic field in two dimensional (2D)
systems. Thus, according to Khmielnitskii \cite{Chmi84} and
Laughlin \cite{Laug84}, extended states associated with Landau
levels should float up and cross the Fermi energy when
$B\rightarrow0$. Hence, the resulting global phase diagram of the
QHE \cite{Kive92} allows only for a direct transition between the
insulating phase and the QHE liquid state with the filing factor
$\nu=1$. If the spin degeneracy is not removed, the diagram
predicts the transition involving  the $\nu=2$ QHE state. However,
a number of recent experiments call into question this picture. In
particular, a zero-field metallic state is observed \cite{Abra01},
the floating scenario is questioned \cite{Krav01}, and transition
from the insulating phase into QHE states with $\nu\geq3$ are
reported \cite{Song97,Lee98,Hilk00}. The latter, however, are
interpreted also as a cross-over, not the quantum phase transition,
occurring between the QHE and weak localization regimes, each
characterized by rather different conductivities at non-zero
temperatures \cite{Bodo00}. We have carried out a detail study of
the boundary between the electron and QHE liquids with the hope
that a large ratio of the spin and cyclotron splittings will result
in a better resolution. Indeed, we detect an additional plateau
$\sigma_{xy} \approx e^2/h$ in a sample with the lowest electron
concentration, which we assign to the floating $\nu =1$ Landau
level.

\begin{figure}[t]
\centerline{\epsfig{file=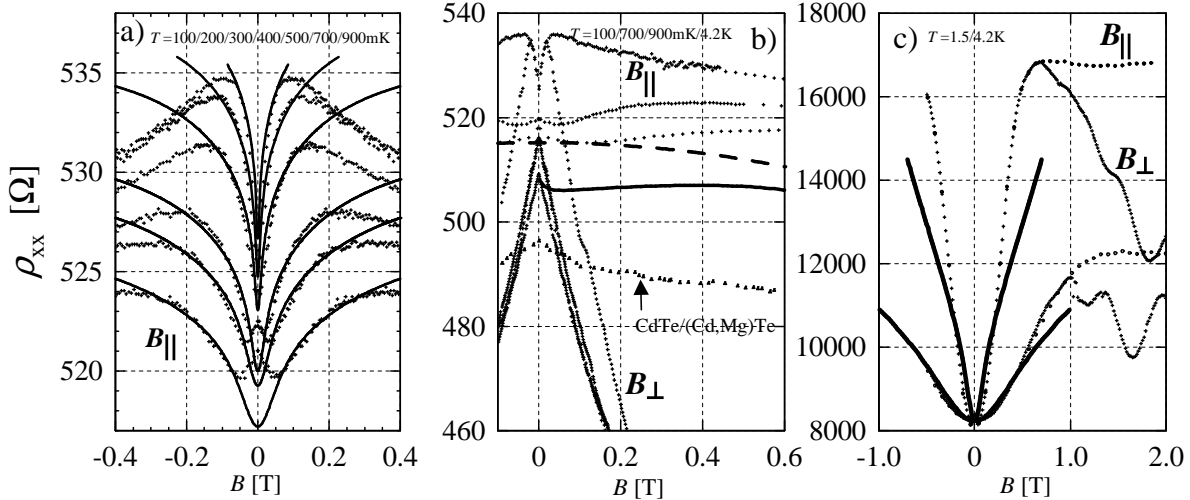,width=16cm,clip=}}
\caption{Low-field magnetoresistance (MR) in  heterostructures
with electron density and net Mn concentration $n_s= 5.9\times
10^{11}$~cm$^{-2}$, $x=0.005$, respectively, in parallel (a) and
perpendicular (b) configurations. (c) MR for $n_s= 1.35\times
10^{11}$~cm$^{-2}$; $x=0.05$.  MR for the nonmagnetic CdTe with
similar density and mobility is also shown in (b). Thick solid
lines are calculated taking effects of disorder on
electron-electron interactions into account
\cite{Alts85,Fuku85,Lee85,Paal83}. The data demonstrate that
current theory explains the positive MR but fails to describe an
additional Mn-induced negative MR in the perpendicular
configuation.}

\end{figure}

The Mn ions were introduced to the quantum wells (QW) either
digitally or uniformly, the mean Mn content varying between
$x=0.005$ and $x=0.05$. In these structures  the electron density
varies between $n_s= 1.35 - 5.9\times 10^{11}$~cm$^{-2}$, and
electron mobility up to 60~000 cm$^2$/Vs is observed. As shown in
Fig.~1, the presence of Mn leads to giant spin-related positive as
well as to negative MR, the latter of an orbital origin as it is
observed only for the magnetic field perpendicular to the QW plane.
The quantitative analysis of the data is based on theory of quantum
corrections to conductivity $\sigma$ of disordered systems
\cite{Alts85,Fuku85,Lee85,Paal83}, using codes and procedures
developed previously \cite{Jaro95,Sawi86,Brun96}. As shown, the
model does not account for the negative MR, which suggests the
presence of a new Mn-induced phenomenon. By contrast, in agreement
with with the earlier studies \cite{Jaro95,Smor97,Sawi86}, we find
that the positive MR stems from the effect of giant spin-splitting
upon interference of electron-electron interaction amplitudes in
disordered systems. The calculated MR agrees well with the
experimental data, particularly for the sample with $n_s= 5.9\times
10^{11}$~cm$^{-2}$. In the case of the sample with the lower
density $n_s= 1.35 \times 10^{11}$~cm$^{-2}$, a discrepancy between
the observed and calculated MR appears at low temperature,
presumably because of the proximity to the strongly localized
regime as well as the effect of the redistribution of the carriers
between the spin subbands, a MR mechanism \cite{Fuku79} neglected
in our computations.

Our results imply that quantum localization effects in 2D systems
make $\sigma$ to be sensitive to spin polarization not only in low
mobility materials studied previously
\cite{Jaro95,Smor97,Sama98,Sawi86} but also in the case of 2D
systems containing high mobility carriers. Actually, the value of
$\sigma$ depends on the ratio of temperature to the energy distance
between quasi Fermi levels for the two spin subbands, and not not
on $\sigma$ itself. This means, in particular, that temperature
dependence of $\sigma$ may serve to determine quantitatively the
efficiency of spin injection into a 2D electron liquid.

\begin{figure}[t]
\centerline{\epsfig{file=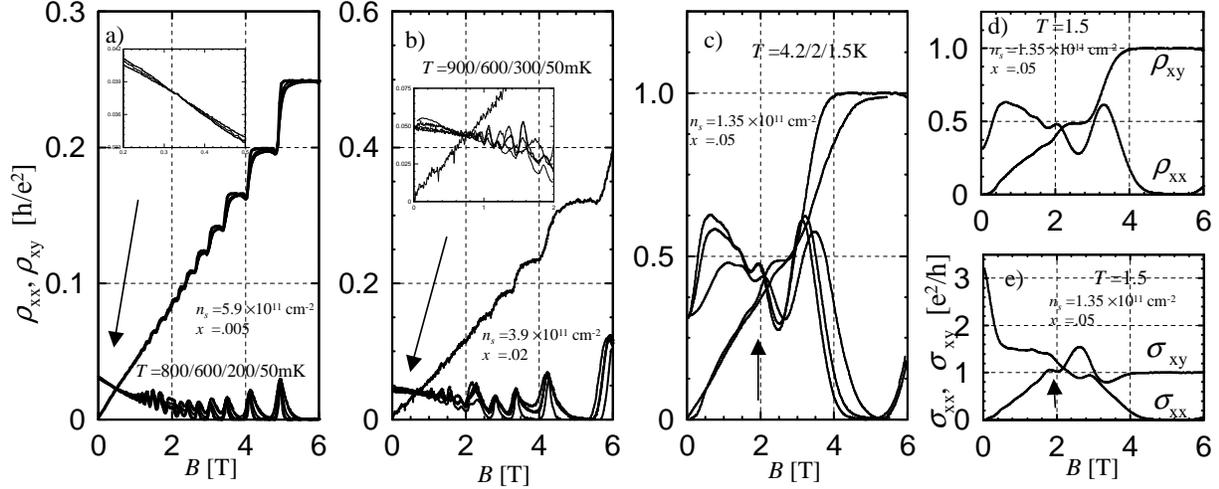,width=16cm,clip=}}
\caption{ Longitudinal and Hall resistivities in devices with (a)
$n_s= 5.9\times 10^{11}$~cm$^{-2}$, $x=0.005$, (b) $n_s= 3.9\times
10^{11}$~cm$^{-2}$, $x=0.02$ and (c) $n_s= 1.35\times
10^{11}$~cm$^{-2}$, $x=0.05$. Insets to (a) and (b) show the the
field range corresponding to the transition between the
spin-polarized liquid and quantum Hall effect regimes. (d)(e)
transport coefficients in the device with $n_s= 1.35\times
10^{11}$~cm$^{-2}$, $x=0.05$ at 1.5~K. Arrow indicates an
additional plateau of $\sigma_{xy}$, assigned to the floating
Landau level.}
\end{figure}

Figure~2 summarizes results of MR and Hall effect measurements for
three samples with increasing disorder and Mn concentrations in the
range of the magnetic field corresponding to the transition between
the classically weak and strong magnetic fields, $\rho_{xx}\approx
\rho_{xy}$. At first glance, the results look similar to those
reported for 2D nonmagnetic systems \cite{Song97,Lee98,Hilk00}. In
particular,  at $\rho_{xx}=\rho_{xy}$,  $\rho_{xx}$ becomes
virtually temperature independent. Furthermore, a cross-over from
the weakly localized regime  into $\nu=11,6,2$ QHE states is
observed for heterostructures with $n_s= 5.9,\,3.9,\mbox{ and }
1.35\times 10^{11} $~cm$^{-2}$, respectively.

However, in the device with density $n_s= 1.35 \times
10^{11}$~cm$^{-2}$ the presence of the Shubnikov-de Haas
oscillations in the magnetic field lower than the cross-over point
are clearly seen. Moreover, an additional
$\sigma_{xy}\approx$~e$^2$/h plateau is observed in this field
range, as shown in Fig.~2(e). We assign this plateau to the
floating lowest Landau level, as theoretically predicted
\cite{Chmi84,Laug84} but hardly observed so far. Most probably, the
large spin-splitting and the enhanced localization have made our
search successful. Interestingly, recent theoretical \cite{Kosc01}
and experimental results \cite{Krav01} indicate  that, at least in
the case of the spin-polarized electron liquid, the extended states
float-up in vanishing magnetic fields.

In conclusion, the spin polarization affects interference of
electron-electron interaction amplitudes, which leads to a large
reduction of conductivity even for a small magnitude of spin
polarization. Under such conditions, a transition between spin
polarized carrier liquid and QHE states has been detected. This
observation corroborates the scenario of floating up Landau levels.
The presence of Mn results also in a large negative orbital
magnetoresistance, whose origin is unknown at present.

This work was supported by Foundation for Polish Science and KBN
grant No.~2--P03B--02417.

\myskip

\end{document}